\documentclass[a4paper]{article} 
\usepackage[margin=1in]{geometry}
\usepackage{graphicx}
\usepackage{hyperref}
\hypersetup{colorlinks=true, linkcolor=blue, filecolor=magenta, urlcolor=cyan}
\usepackage{color}
\usepackage{soul}
\usepackage{authblk}
\newcommand{\myindent}{\hspace{1cm}}
\usepackage[labelfont=bf]{caption}

\begin{document}

\title{Extended X-ray absorption spectroscopy using an ultrashort pulse laboratory-scale laser-plasma accelerator}

\newcommand{\JAI}{The John Adams Institute for Accelerator Science, Imperial College London, London, SW7 2AZ, UK}
\newcommand{\QUB}{School of Mathematics and Physics, Queen's University of Belfast, BT7 1NN, Belfast, UK}
\newcommand{\CUOS}{Gérard Mourou Center for Ultrafast Optical Science, University of Michigan, Ann Arbor, Michigan 48109-2099, USA}
\newcommand{\Lund}{Department of Physics, Lund University, P.O. Box 118, S-22100, Lund, Sweden}
\newcommand{\CLF}{Central Laser Facility, STFC Rutherford Appleton Laboratory, Didcot OX11 0QX, UK}
\newcommand{\HZDR}{Helmholtz-Zentrum Dresden-Rossendorf, Bautzner Landstrasse 400, 01328 Dresden, Germany}
\newcommand{\ASCR}{Institute of Physics of the ASCR, Na Slovance 1999/2, 182 21 Prague, Czech Republic}
\newcommand{\TUD}{Technische Universit{\"a}t Dresden, D-01069 Dresden, Germany}
\newcommand{\LLNL}{Lawrence Livermore National Laboratory (LLNL), Livermore, California 94550, USA}
\newcommand{\PULSE}{Stanford PULSE Institute, SLAC National Accelerator Laboratory, Menlo Park, California 94025, USA}
\newcommand{\JM}{Johnson Matthey Technology Centre, Blount's Court, Sonning Common, Reading, RG4 9NH, United Kingdom}

\author[1]{Brendan Kettle*}
\author[1]{Cary Colgan}
\author[1]{Eva E.~Los}
\author[1,2]{Elias Gerstmayr}
\author[2]{Matthew J.~V.~Streeter}
\author[3]{Felicie Albert}
\author[4]{Sam Astbury}
\author[1]{Rory A.~Baggott}
\author[2]{Niall Cavanagh}
\author[5,6,7]{Kate\v{r}ina Falk}
\author[8]{Timothy I.~Hyde}
\author[9]{Olle Lundh}
\author[4]{Rajeev P.~Pattathil}
\author[2]{Dave Riley}
\author[1]{Steven J.~Rose}
\author[2]{Gianluca Sarri}
\author[4]{Chris Spindloe}
\author[9]{Kristoffer Svendsen}
\author[4]{Dan R.~Symes}
\author[5]{Michal \v{S}m\'{i}d}
\author[10]{Alec G.~R.~Thomas}
\author[4]{Chris Thornton}
\author[1,11]{Robbie Watt}
\author[1]{Stuart P.~D.~Mangles}

\affil[1]{\small{The John Adams Institute for Accelerator Science, Imperial College London, London, SW7 2AZ, UK}}
\affil[2]{School of Mathematics and Physics, Queen's University of Belfast, BT7 1NN, Belfast, UK}
\affil[3]{Lawrence Livermore National Laboratory (LLNL), Livermore, California 94550, USA}
\affil[4]{Central Laser Facility, STFC Rutherford Appleton Laboratory, Didcot OX11 0QX, UK}
\affil[5]{Helmholtz-Zentrum Dresden-Rossendorf, Bautzner Landstrasse 400, 01328 Dresden, Germany}
\affil[6]{Institute of Physics of the ASCR, Na Slovance 1999/2, 182 21 Prague, Czech Republic}
\affil[7]{Technische Universit{\"a}t Dresden, D-01069 Dresden, Germany}
\affil[8]{Johnson Matthey Technology Centre, Blount's Court, Sonning Common, Reading, RG4 9NH, United Kingdom}
\affil[9]{Department of Physics, Lund University, P.O. Box 118, S-22100, Lund, Sweden}
\affil[10]{Gérard Mourou Center for Ultrafast Optical Science, University of Michigan, Ann Arbor, Michigan 48109-2099, USA}
\affil[11]{Stanford PULSE Institute, SLAC National Accelerator Laboratory, Menlo Park, California 94025, USA}

\date{} 

\maketitle

\vspace{-0.5cm}
*Corresponding author email: b.kettle@imperial.ac.uk



\section*{Abstract}

Laser-driven compact particle accelerators can provide ultrashort pulses of broadband X-rays, well suited for undertaking X-ray absorption spectroscopy measurements on a femtosecond timescale.
Here the Extended X-ray Absorption Fine Structure (EXAFS) features of the K-edge of a copper sample have been observed over a 250 eV window in a single shot using a laser wakefield accelerator, providing information on both the electronic and ionic structure simultaneously. 
This unique capability will allow the investigation of ultrafast processes, and in particular, probing high-energy-density matter and physics far-from-equilibrium where the sample refresh rate is slow and shot number is limited. 
For example, states that replicate the tremendous pressures and temperatures of planetary bodies or the conditions inside nuclear fusion reactions. 
Using high-power lasers to pump these samples also has the advantage of being inherently synchronised to the laser-driven X-ray probe. 
A perspective on the additional strengths of a laboratory-based ultrafast X-ray absorption source is presented. 



\section*{Introduction}

In recent years laser-plasma based particle accelerators have provided access to gigaelectronvolt electron energies within the small-scale laboratory environment \cite{Esarey2009,Albert2021,Gonsalves2019}.
This has led to new research involving strong field QED studies~\cite{Cole2018,Poder2018}, electron-positron pair generation~\cite{Kettle2021,Sarri2015} and X-ray and gamma ray applications~\cite{Albert2014,Cole2015,Sarri2014,Schumaker2014}.
One growing area makes use of the X-rays generated in tandem with the accelerated beam as the electrons wiggle in the back of the plasma wakefield bubble~\cite{Kneip2010,Corde2013}.
The X-rays have a pulse duration comparable to the emitted electron bunch; usually 10's of femtoseconds~\cite{Debus2010, Lundh2011,Mangles2006}.

\myindent A key strength of this X-ray source is that it has a smooth, broadband synchrotron-like spectrum, making it ideal for X-ray absorption spectroscopy (XAS) techniques such as X-ray Absorption Near-Edge Structure (XANES) and Extended X-ray Absorption Fine Structure (EXAFS) spectroscopy.
In these techniques the absorption, scattering and interference of ejected photoelectrons from neighboring atoms manifest as modulations in the absorption profile near resonant edges. 
These modulations are directly linked to the local electronic and ionic structure of the sample~\cite{Rehr2000,KoningsbergerPrins1988}. 
In fact the electron temperature, ion temperature, ionisation state and local ionic positions can be simultaneously measured~\cite{Dorchies2016}. 

\myindent A major application lies in the study of extreme states far from equilibrium, or those with high energy density, such as the tremendous pressures and temperatures of planetary formations~\cite{Koenig2004}, or  inertial confinement fusion experiments~\cite{Betti2016}.
These conditions are complex to understand as they can demonstrate strong ion coupling, both long- and short-range order with both free and bound electrons becoming strongly correlated, or exhibit partial ionisation as well as quantum effects such as degeneracy.
Measuring the electronic and ionic structure of these samples experimentally is key to understanding the states further.
For example, nanosecond EXAFS has been performed on high-energy laser systems to investigate materials at high pressures~\cite{Yaakobi2004, Ping2013PRL, Chin2023}.
Many processes under investigation however, such as the electron-ion equilibration rate~\cite{Mahieu2018,Grolleau2021}, non-thermal phase changes~\cite{Rousse2001}, or bond-hardening effects~\cite{Ernstorfer2009}, require a probe of ultrashort duration to capture the transient dynamics. 
For these experiments a single-shot measurement is also of major benefit, as driving the sample to the appropriate conditions requires extensive resources and shot rate is limited, with the sample being destroyed each time.

\myindent Until recently, the photon flux from laser-plasma accelerator based XAS measurements has required 10's to 100's of shots to form an absorption profile, which often suffers from noise and is limited to the near-edge structure~\cite{TaPhuoc2007, Mahieu2018}.
Recently single-shot measurements using a laser-plasma accelerator have been demonstrated~\cite{Kettle2019}, however the spectral range was limited to 40~eV, and so only XANES features were visible, i.e. the ion structure, was unavailable.
Here, using a new experimental geometry, we demonstrate a pivotal increase in signal-over-noise and extend the single-shot spectral range to over 250 eV.
This allows access to the EXAFS region for the first time, a significant capability for performing pump-probe experiments of high-energy-density and non-equilibrated states.
The potential of ultrafast laser-plasma XAS experiments is furthered by the fact the source can be co-located with other high-power lasers for pumping samples, and can be inherently synchronised on a femtosecond level.

\myindent In this article we present a platform for a laser-plasma accelerator based extended X-ray absorption spectroscopy system that will enable single-shot measurements, of particular importance for targets driven to extreme states.
We will also discuss the additional advantages of providing a technique with the unique capability of measuring ultrafast processes on a femtosecond timescale within a laboratory-scale environment.


\section*{Results}

\begin{figure*}[!t]
\centering
\includegraphics[width=0.95\textwidth]{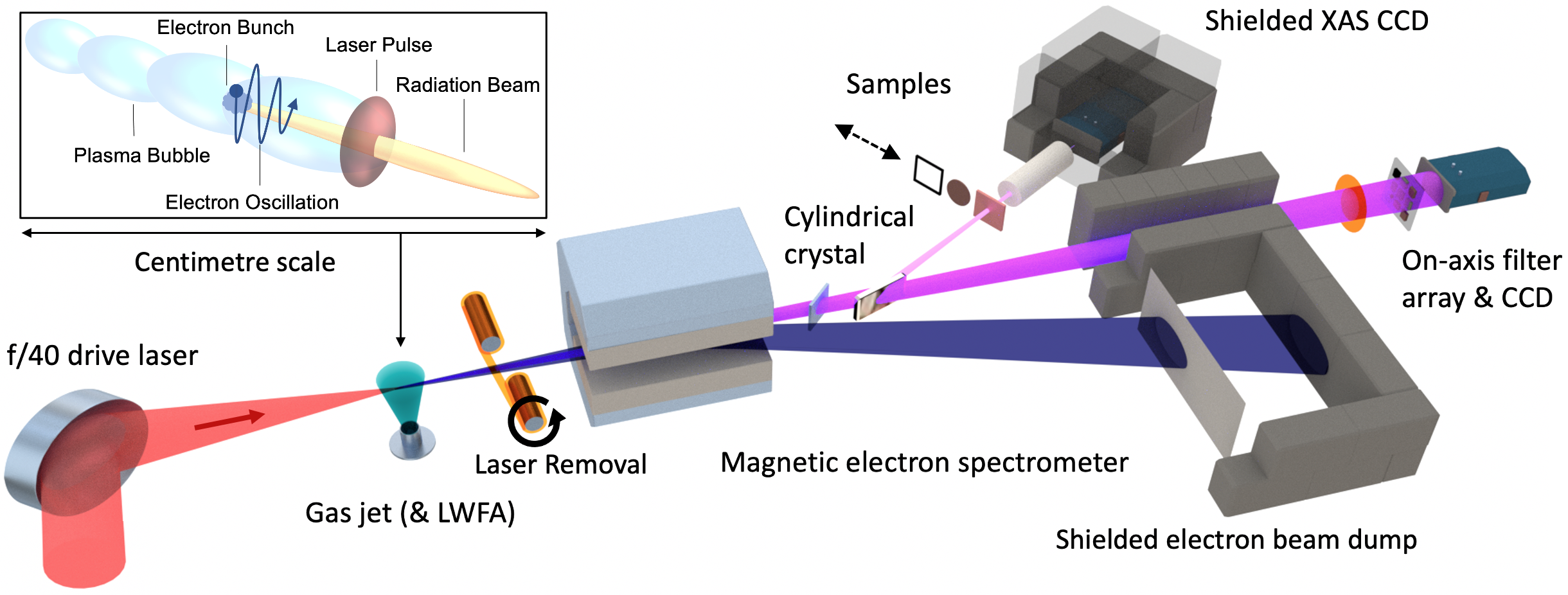}
\caption{\label{fig:setup} \textbf{Overview of the instrument setup.} An ultrashort laser pulse creates a laser wakefield accelerator (inset) in a high pressure gas jet, generating a beam of on-axis electrons and X-rays. 
A refreshable tape drive removes the remaining drive laser light from the beam path.
The electrons are swept away from the beam axis using a dipole magnet and spectrally diagnosed using a scintillator screen before being dumped in a shielded area.
The X-rays are either diffracted off a cylindrically curved crystal through a sample and spectrally spread onto a shielded CCD, or alternatively (with the crystal translated out) straight onto a CCD through a multi-element filter array used to infer the broad spectral shape.
A thin aluminium filter tilted at $45^o$ is used to protect the X-ray crystal from any debris or stray laser light. }
\end{figure*}

\textbf{Experiment setup.} An overview of the setup can be seen in Fig.~\ref{fig:setup}.
Each drive laser pulse (provided at 0.05 Hz) had a duration of $45\pm5$ fs and contained $6.0\pm0.3$ J, corresponding to a peak laser power of $134\pm18$ TW. 
This provided an on-target intensity of ($3.0\pm0.4)\times10^{18}~\textrm{W~cm}^{-2}$ and a laser strength parameter of $a_0=1.2\pm0.1$. 
As this pulse propagates through the gas, it drives a laser wakefield accelerator (LWFA), expelling electrons from the atoms and creating a charge cavity in the wake of the laser pulse. 
The electrons are accelerated to relativistic energies in this wake, emitting X-ray synchrotron light as they oscillate at the back of the cavity.
The electron beam is subsequently dumped into a lead shielded cavity, while the broadband X-ray beam is reflected off a cylindrically curved HAPG (Highly Annealed Pyrolytic Graphite) crystal, onto a well shielded X-ray CCD.
The crystal spectrally disperses the X-rays in one axis, while focusing them spatially in the other axis.
Various copper samples were placed on a translation stage before the CCD, so that their absorption could be measured.
By temporarily moving the crystal, it was also possible to infer the full broadband spectral shape of the X-rays using an on-axis CCD with an elemental filter array placed in front~\cite{Kneip2008, Kneip2010, King2019}.\\

\begin{figure}[!h]
\centering
\includegraphics[width=0.75\textwidth]{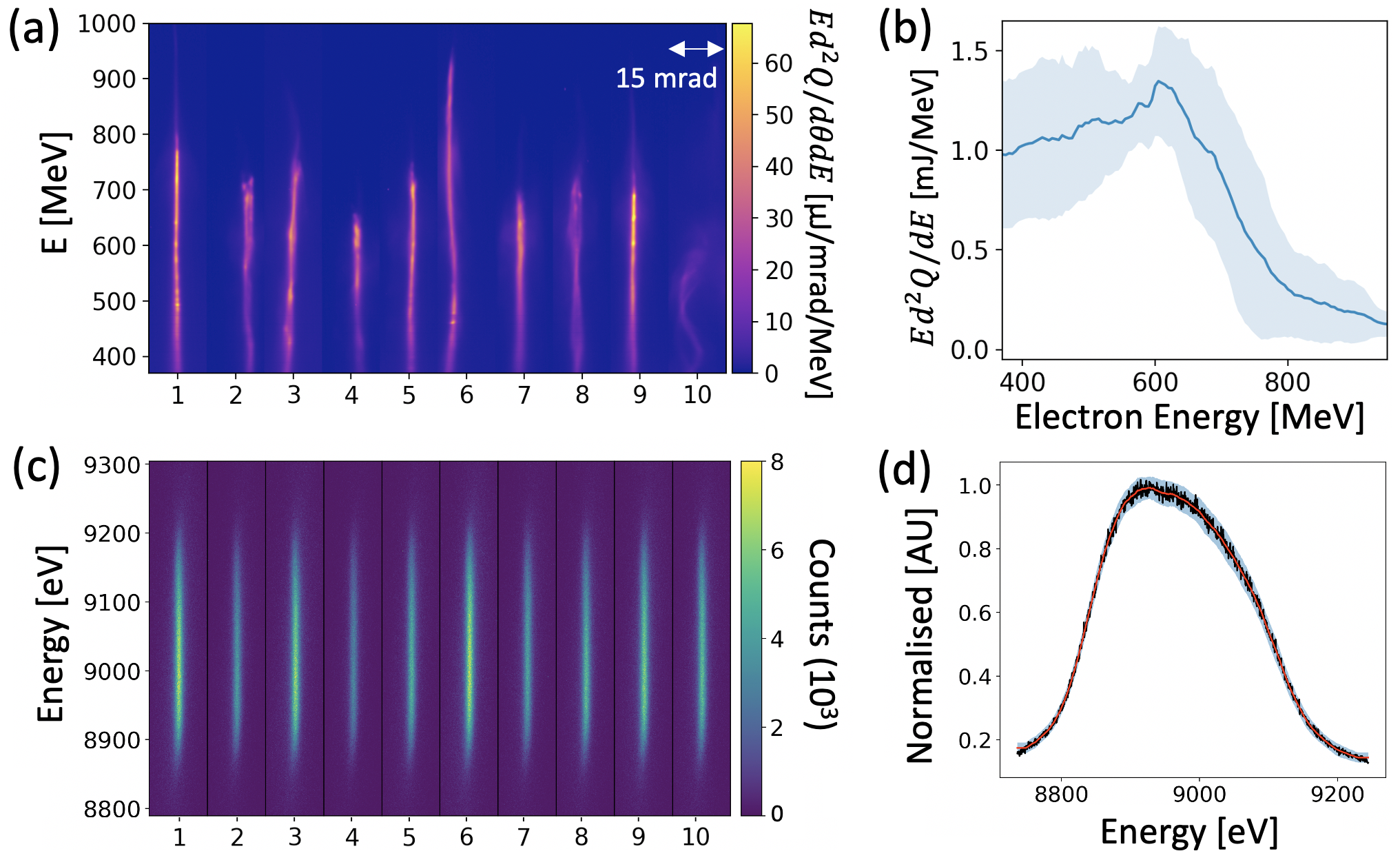}
\caption{\label{fig:source} \textbf{Example electron and X-ray data.} (a) Electron spectra for ten consecutive shots. (b) Mean and standard deviation of these spectra (shaded). (c) Raw XAS CCD data for the same shots. The X-rays are spatially focused to a line, and spectrally spread in the vertical axis. (d) Smoothed X-ray reference profile shape (red), the unsmoothed average of the 10 normalised profiles (black), and the standard deviation (shaded blue); the X-rays on the spectrometer demonstrate a stable and consistent spectral shape despite the variation in electron beam properties.}
\end{figure}

\textbf{Electron and X-ray source properties.} For the data presented here, the LWFA was tuned to maximise the generated photon beam flux in our desired spectral range, which coincided with a relatively high-charge broadband electron beam~\cite{Kettle2019}. 
This was primarily achieved by increasing the density of the gas jet. 
Fig.~\ref{fig:source} (a) and (b) depicts the electron spectra for 10 consecutive example shots.
The 90th percentile electron energy was $680\pm40$ MeV, with a charge of $100\pm30$ pC per shot.
The spectral shape of the broadband on-axis X-ray beam was assumed to be synchrotron-like, and characterised by a critical energy, $E_{\mathrm{crit}}$ (see methods). 
An average critical energy of $E_{\rm{crit}} = 19.6 \pm 1.0~\mathrm{keV}$, and a standard deviation of $\sigma_{\mathrm{E_{crit}}}=3.4~\mathrm{keV}$ was found (inferred over 10 shots in the run prior to the data of Fig.~\ref{fig:source}).
Fig.~\ref{fig:source} (c) depicts the X-ray signal for the same 10 shots of Fig.~\ref{fig:source} (a).
On average $(2.4\pm0.5)\times10^5$ photons per eV were emitted in the 9~keV region per shot.
It is important to note that while the electron beam exhibits shot-to-shot variability, with the 90th percentile electron beam energy varying by 150 MeV (22\%) across the run and a shot-to-shot variation in total charge of 29\%, the measured portion of 9 keV radiation is seen to be spectrally smooth, and can be easily scaled in magnitude.
This is because the crystal spectrometer observes a small slice of the broadband synchrotron spectrum, a relatively slowly varying function when concerned with the fluctuations in total electron beam energies we observe. 
By averaging over the profiles of the X-ray data in Fig.~\ref{fig:source} (c), a reference spectral profile shape is found (each single shot is normalised to its peak level, and a rolling average smoothing function is applied).
This profile is given in Fig.~\ref{fig:source} (d).
The variation from the measured signal is less than a few percent and so it is possible for the absorption profile of a sample to be measured without an on-shot reference (but rather an assumed profile).
In summary, the stability of the X-ray source for absorption measurements is evident, despite the fluctuations of the electron beam properties.\\

\begin{figure}
\centering
\includegraphics[width=0.6\textwidth]{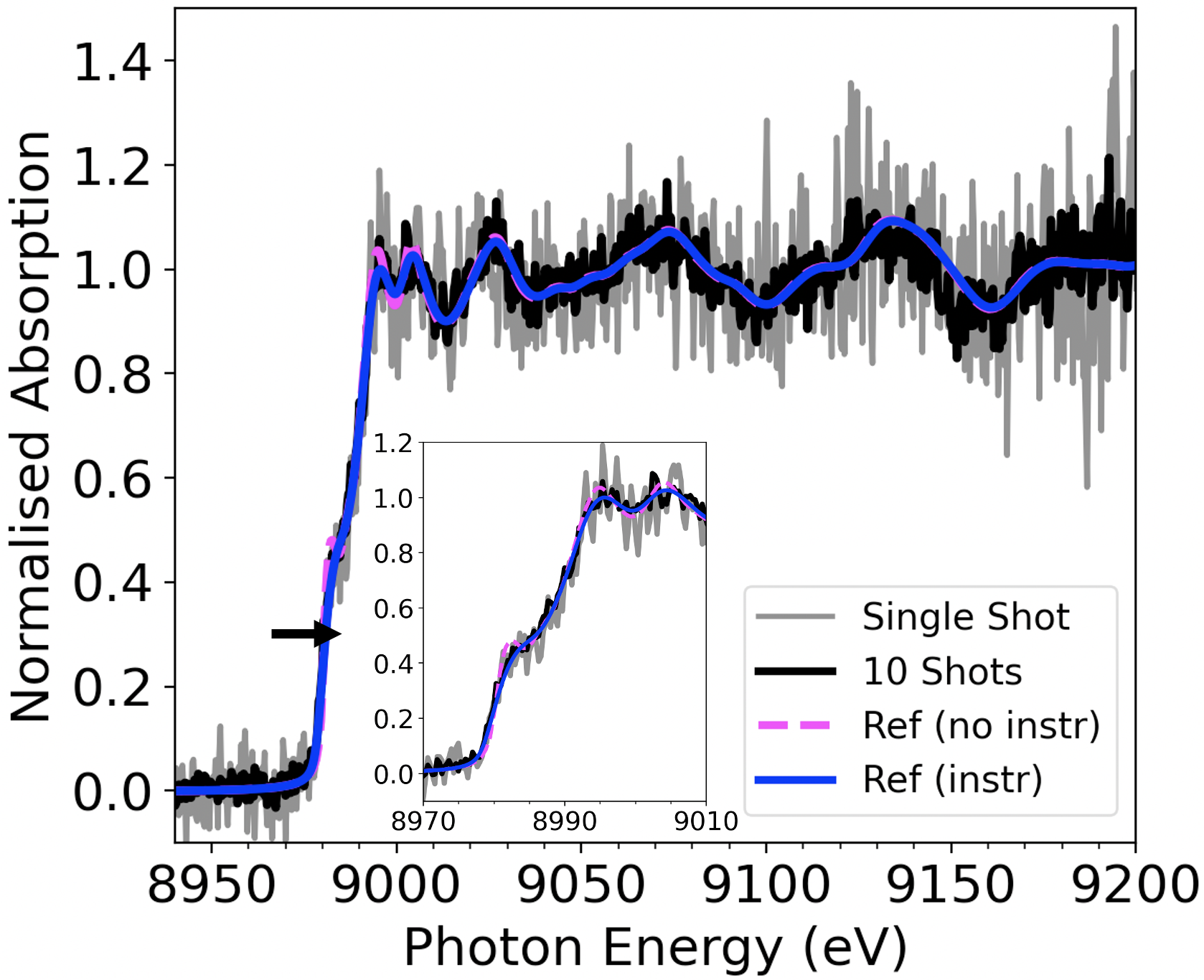}
\caption{\label{fig:abs_results} \textbf{Normalised absorption profile for a copper foil.} Data is shown for a single shot (grey) and an average of ten shots (black). Both are compared to a synchrotron reference convolved with (blue) and without (magenta) our simulated instrument response. Inset: zoomed view of the XANES profile near the edge.}
\end{figure}

\textbf{X-ray absorption results.} The absorption profile of a $4~\mathrm{\mu m}$ copper sample has been measured, assuming the reference profile of Fig.~\ref{fig:source} (d) as the unattenuated signal (the known transmission before the copper K-edge is used to scale the magnitude).
Fig.~\ref{fig:abs_results} depicts the resulting normalised absorption data for the copper sample.
The inset gives the XANES profile close to the edge.
The profile for a single shot (grey) and the average profile for a 10 shot run (black) is shown.
Both are compared to the absorption profile of a similar copper foil sample measured at B18, the general purpose XAS beamline at the Diamond Light Source synchrotron facility.
This reference is presented both with the simulated instrument function of our system applied (blue solid), and the unaltered synchrotron data (dashed magenta).
Our instrument response was simulated using the MMPXRT code~\cite{Smid2021} and  indicates a spectral resolution of $\approx5$ eV (FWHM).
Note that the reference data without our instrument function applied differs only slightly at the edge (inset).
The measured profile stretches over a 250 eV spectral range, allowing access to the EXAFS region of the profile.
It demonstrates a significant increase in signal-to-noise over previous measurements~\cite{Kettle2019}, as well as overall photon number from other LWFA X-ray absorption measurements~\cite{Mahieu2018}.
The two most important alterations to the experimental setup were to use a spatially focusing cylindrically curved crystal, and to mitigate the noise on the X-ray detector from the electron beam.
Both these factors contributed greatly to an increased signal-to-noise ratio. 
The noise mitigation was designed after consulting various FLUKA simulations~\cite{Fluka2014}, where it was deemed it was most important to sweep the electron beam in the opposite direction to the X-ray deflection, and into a well shielded beam dump, with the X-ray CCD separately enclosed in its own shielding, that includes a shielded tunnel from the CCD in the direction of the crystal.
In fact, we infer that the noise level on a single shot is close to Poisson limited (where for peak signal we have $\approx500$ photons per spectral pixel, with a standard deviation of $\approx23$ photons, or $\approx\sqrt{N_{\mathrm{ph}}}$).
This indicates we have removed the majority of any environmental noise sources, and are limited by the measured photon number of our signal. 
As the X-ray source has a divergence on the milliradian scale, it should be relatively easy to decouple the environmental noise due the electron beam when scaling up to larger platforms such as high-energy-density systems, by using appropriate magnetic deflection and introducing distance from the electron beam dump and the shielded X-ray detector.
When considering the addition of a pumped target, the narrow beamed nature of the X-ray probe is also an advantage when trying to mitigate any self-emission from samples as this emission  will radiate into a $4\pi$ sphere, with an inverse distance-squared dependence. 
The X-ray source can also be collimated or refocused with the correct optics. 
Additional optics after the sample can also re-direct the X-ray probe, removing any photons outside the spectral window of interest. 
Similarly, the correct choice of filtering in front of the detector will help mitigate any sample noise that lies in a different region of the spectrum. \\

\begin{figure}[!h]
\centering
\includegraphics[width=0.7\textwidth]{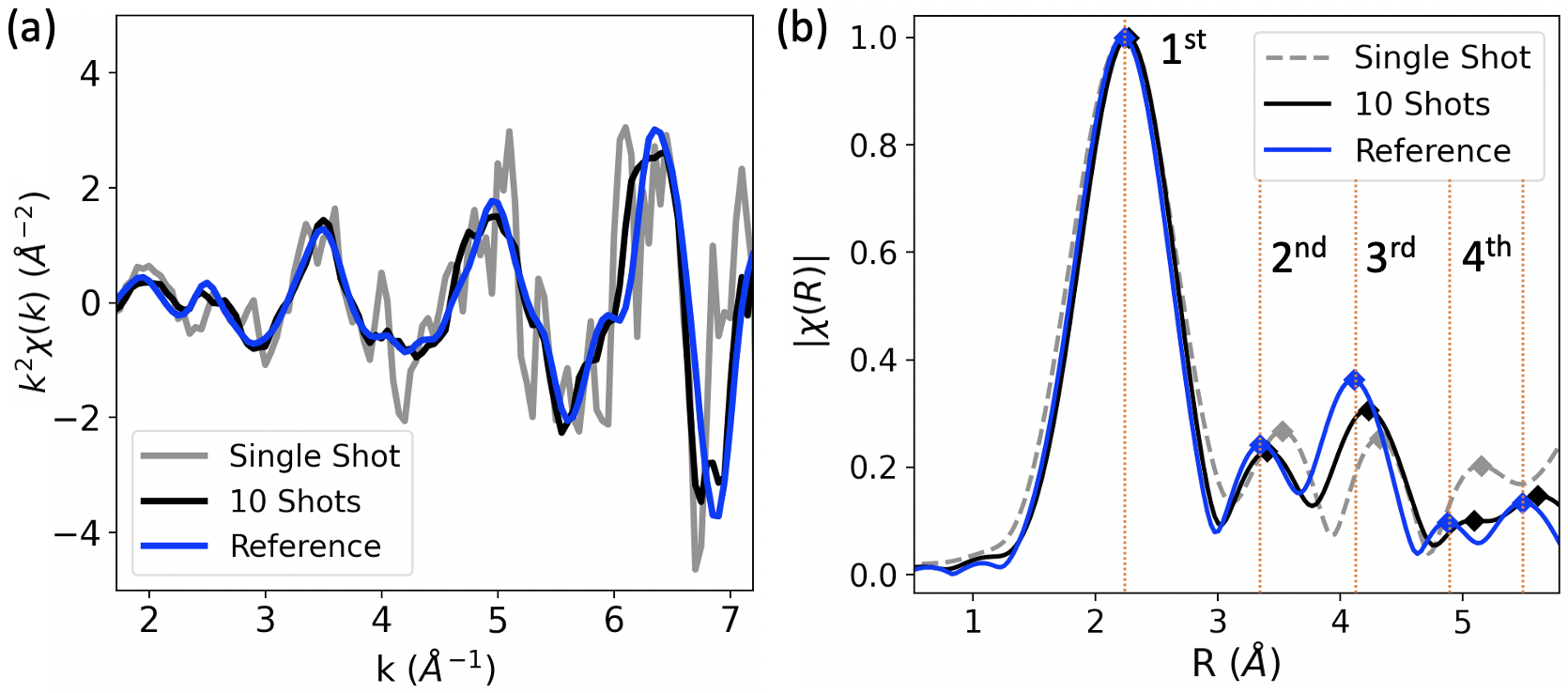}
\caption{\label{fig:exafs_results}\textbf{Extended fine structure results.} (a) Isolated EXAFS oscillations in $k$-space (b) The Fourier transform to $R$-space, where peaks correspond to the distance to neighbouring atoms, or coordination shells (labelled). Data is shown for a single shot (grey) and an average of ten shots (black). Both are compared to a synchrotron reference convolved with our simulated instrument response (blue).}
\end{figure}

\textbf{Extended X-ray absorption.} In Fig.~\ref{fig:abs_results} the EXAFS profile after the edge is visible on a single shot, to the best of our knowledge, a first of its kind measurement using a femtosecond probe. 
The processed EXAFS features are presented in Fig.~\ref{fig:exafs_results}.
Details on the processing are found in the methods sections.
Fig.~\ref{fig:exafs_results} (b) compares our data to the synchrotron reference data, processed under the same conditions.
$|\chi(R)|$ is similar to a radial distribution function, and the peaks correspond to the distance to neighboring atoms (or coordination shells).
The scattering from the first four shells are labelled~\cite{Zabinsky1995}.
Our single shot data matches the first shell of the reference data to within 1.5\%, and shells 2-4 within 5\% (a peak finding algorithm was used to determine their location).
The 10 shot data improves on this, with the second, third and fourth shells matching within 2\%, 3\% and 5\% respectively. 
This is a direct measurement of the local atomic structure, and unique to the species of a sample.
Interpretation of this structure also allows access to information on the ion temperature, any sample compression or phase changes; important quantities for understanding the properties of high-energy-density and non-equilibirium states. 

\section*{Discussion}

The signal-to-noise ratio we have achieved has made it possible to produce high quality measurements of both the XANES and EXAFS features in a single shot using a femtosecond probe.
This enables probing of ultrafast processes, but especially those samples in extreme conditions which are limited to a low shot rate, and samples are damaged each measurement.
The development of this ultrafast XAS platform is on-going, with clear routes for further improvements on flux, photon energy and spectral resolution.

\myindent Petawatt-class laser systems are becoming more commonplace, providing nearly an order of magnitude more peak laser power than the experiment we detail.
These systems can access higher electron energies~\cite{Gonsalves2019,Thomas2010}, and increased X-ray flux~\cite{Thomas2010}.
Scaling of the spectral shape (characterised by $E_{\mathrm{crit}}$, the critical energy) and the peak X-ray brightness $B_0$, with drive laser power and plasma density are described in Kneip \textit{et al.}~\cite{Kneip2010HEDP}.
Using the relation in Mangles \textit{et al.}~\cite{Mangles2012} to estimate the  density at which self-injection occurs in terms of peak laser power, we can express the scalings in terms of drive laser power alone~\cite{Bloom2020}. 
The critical energy of the synchrotron-like spectrum $E_{\mathrm{crit}}$ is seen to scale approximately linearly with peak laser power P, and the peak brightness $B_0$ is seen to scale approximately as $\propto P^{1.6}$.
The measurements presented here were made using $\approx$130 TW of laser power, thus moving to 1~PW, one would expect an increase of approximately $\times26$ in peak X-ray brightness.
The critical energy of the emitted spectrum also increases from $\approx20$ keV to $\approx150$ keV. 
This is particularly important for studying samples above 30 keV, as these photon energies cannot be accessed easily by conventional synchrotron facilities, and will provide the capability to probe higher-Z elements and inner atomic edges, for example, in nuclear research~\cite{Lind2007, Hojbota2022}.
Accounting for the shift in spectral shape, for the photon energy region accessed in this article ($\approx 9$ keV), moving to 1 PW results in approximately $\times9$ more X-ray flux.

\begin{figure}[!h]
\centering
\includegraphics[width=0.8\textwidth]{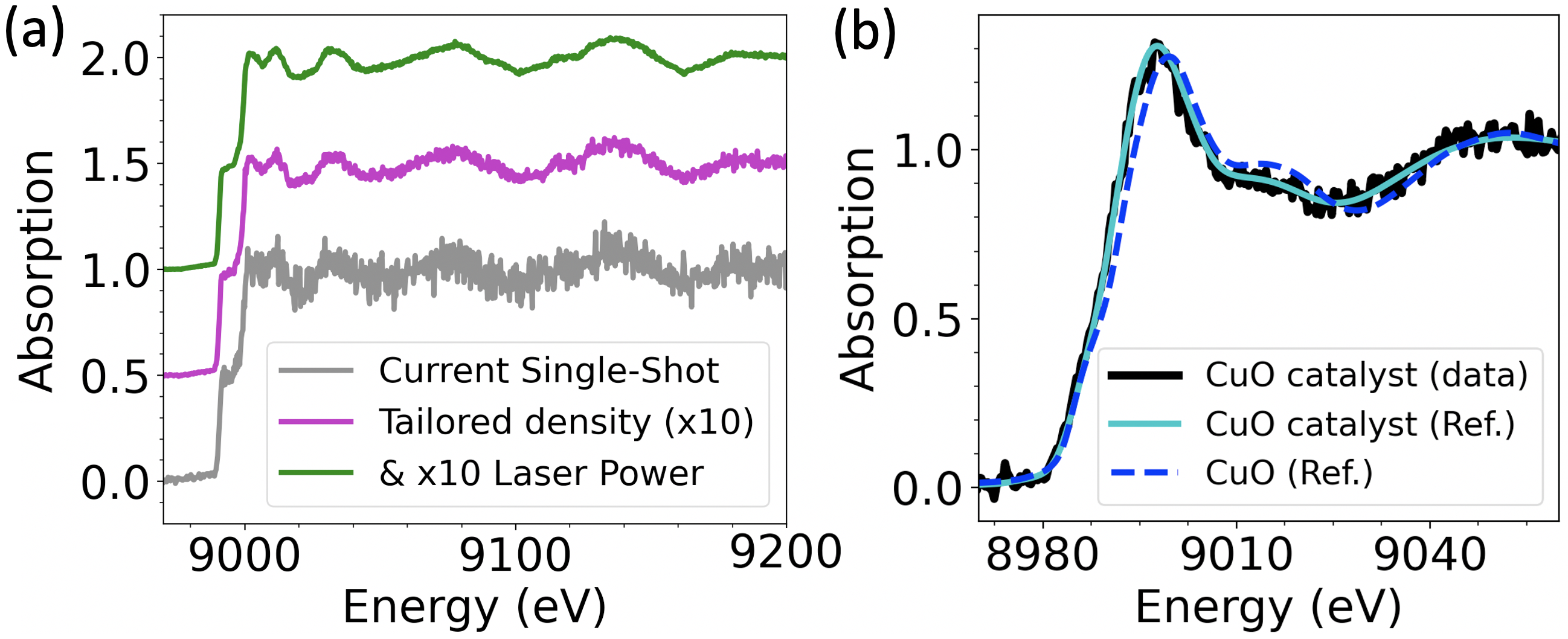}
\caption{\label{fig:perspective} \textbf{Signal predictions and example applications.} (a) Predictions for signal quality with tailored gas targetry and increased laser power (plots have been shifted vertically for clarity). (b) Example absorption profile for a copper oxide catalyst (black), compared to a synchrotron scan (cyan) and a standard copper oxide sample (blue dashed).}
\end{figure}

\myindent Tailored targetry techniques can also be used to increase the source X-ray flux.
Recent results have shown that introducing a density gradient into the target, using a wire or razor blade obstruction, can increase the electron oscillation radius and frequency, enhancing the X-ray emission by an order of magnitude~\cite{Rakowski2022, Kozlova2020, TaPhuoc2008}. 
By combining the above improvements in laser technology and targetry one can speculate two orders of magnitude improvement in signal-to-noise ratio. 
Fig.~\ref{fig:perspective} (a) depicts a simulated comparison of these improvements, based on the noise being Poisson-limited by the photon number.
With a $\times10$ increase in flux provided by laser power, and a similar increase due to targetry design, we predict that the absorption profile will approach the quality associated with a conventional synchrotron measurement. 
These estimates are conservative, but also given the quality increase in the signal-to-noise, allows for compromise when moving to pump-probe schemes where other factors such as additional X-ray optics would decrease the result signal level.

\myindent Other improvements to the technique are possible. 
The spectral resolution can be increased by using a perfect crystal (such as Si or Ge) instead of a mosaic crystal, if a sacrifice in X-ray flux can be allowed. 
Simulations using MMPXRT~\cite{Smid2021} suggest sub-eV resolution in such a configuration. 
The small source size of the X-rays (sub-micron~\cite{Kneip2010HEDP}) can be harnessed in conjunction with a suitable focusing optic, such as a toroidal crystal, to allow micron focusing of the probe beam. 
This can be used to probe small sample volumes (for example in HED targets) or high-resolution scanning of inhomogeneities in industrial samples such as batteries or spintronic materials.
The stability of the electron accelerator can be improved through laser techniques~\cite{Dopp2017,Pousa2022}, and recent studies have shown that machine learning can be used to optimise the X-ray emission~\cite{Shalloo2020, Ye2022}.

\myindent In addition to the above, petawatt lasers can now operate at repetition rates of 1~Hz, providing an increased rate of data taking ($\times20$ increase from this experiment).
This would allow laser-plasma accelerator based XANES and EXAFS platforms to complement synchrotron systems, i.e. making a similar quality scan in a few minutes.
As an example, using our current source capability, Fig.~\ref{fig:perspective} (b) depicts an ex situ absorption profile of a copper based hydrogenation catalyst provided by Johnson Matthey, UK. 
These samples are of interest for improving conversion rate, lowering energy consumption or tuning the selectivity of chemical reactions.
Our data (black), and the profile for the same sample scanned at the Diamond Light Source (cyan), is compared to a CuO reference sample (dashed blue). 
The characteristic features of this industrial catalyst are clear, indicated by a shifted edge at 8995 eV and an altered near-edge shape at 9015 eV, even with the Cu loading at 13 wt\% (percentage by weight) as expressed as an oxide.
The oxidation state shows the Cu to be oxidic and predominantly in the $\mathrm{Cu}^{2+}$ oxidation state in comparison to the CuO reference sample.
The data shown took 30 shots or 10 minutes (where the scan time at the synchrotron was $\approx$ 3 minutes), but with the improvements discussed in the previous paragraph, a profile with an improved signal-over-noise could be generated in under a minute.
The benefit of a laser-plasma based XAS system as described here is that it could be achieved in a relatively small laboratory environment, in comparison to a conventional synchrotron light source that requires a large national facility.
This reduced scale makes it an accessible tool for single institutions and makes long-term programmatic research possible.
A laboratory scale instrument also allows dedicated access for immediate probing of delicate samples, appropriate handling of hazardous samples (e.g. biological), or analysis of radiological materials for environmental remediation (nuclear waste management)~\cite{Zhong2021, Mottram2020}. 

\myindent It is important to highlight that the primary application of XANES and EXAFS over the last few decades has been using synchrotron facilities to measure the structure of materials, across a wide range of sciences~\cite{SyncRev2018,Bilderback2005}. 
However, when the duration of the X-ray probe is important, measurements are limited to the pulse duration of the synchrotron; generally on the order of hundreds of picoseconds~\cite{Torchio2016}, unless additional slicing techniques are implemented (that sacrifice flux).
X-ray Free Electron Laser (XFEL) facilities operate at femtosecond pulse duration but provide a narrow bandwidth, requiring a scan over many shots to obtain a broad absorption spectrum~\cite{Lemke2013}.
The combined ultrashort and broadband capabilities of a laser-plasma accelerator X-ray source makes it uniquely suited to investigate ultrafast processes on a femtosecond scale through XAS techniques.
These include femtochemistry phenomena, such as photodissociation, spin crossover processes~\cite{Bressler2008, Gawelda2007, Chen2001, Chen2005}, photobiology~\cite{Bressler2010,Jonas1996,Grondelle1994,Arcovito2005}, or photocatalysts and photoelectrodes~\cite{Uemura2017, Hu2021, Park2022}, which are becoming increasingly important, for example in energy storage research. 
Another active area is that of spintronics for data storage and transfer, with specific interest in antiferromagnets~\cite{Hirohata2020, Li2019, Johnson2015, Strungaru2022}.


\section*{Conclusion}

In conclusion, we have demonstrated that the X-rays from a LWFA source are sufficiently bright and stable to measure the absorption profile around the copper K-edge, over a 250 eV spectral window in a single shot.
It was possible to make single-shot EXAFS measurements with a LWFA-based source providing direct access to the local ionic structure of a sample using a femtosecond probe. 
In particular EXAFS is sensitive to short-range order and can provide the ion temperature of a sample, both difficult measurements to make.
Single shot capability of an ultrafast probe is of significant benefit for studying transient pump-probe samples of high-energy-density physics such as warm dense matter and other non-equilibrium states, as well as other ultrafast electronic and ionic processes.
The compact size will also allow the instrument to be co-located with other laboratories more easily, especially XFEL's or high-power laser facilities (where it can be inherently synchronised on a femtosecond level), acting as an auxiliary probe.
In addition, by implementing an XAS source on a laboratory scale, the breadth and depth of both industrial and programmatic scientific research that can be achieved on a wide-scale basis is vastly increased. 


\section*{Methods}

\textbf{Experiment details.} The study was conducted using the Gemini Laser at the Central Laser Facility (U.K.). 
The drive laser (800 nm) was focused using an $f$/40 geometry to a spot of $49\pm2~\mu\textrm{m} \times 52\pm2~\mu\textrm{m}$ FWHM, with the central FWHM containing $44\pm2\%$ of the energy.
The gas jet nozzle was 15 mm diameter, and a 99\% He and 1\% $\mathrm{N}_2$ mix was used to promote ionisation injection of electrons into the charge cavity~\cite{Pak2010,McGuffey2010}.
The nozzle provided a plasma density of $n_e=(2.1\pm0.2)\times10^{18}~\mathrm{cm}^{-3}$.
After the LWFA, the residual drive laser was blocked by a refreshable tape drive (125 $\rm{\mu m}$ Kapton), and the electron beam was diagnosed with a \mbox{0.3 $\mathrm{T}\,\mathrm{m}$} magnetic spectrometer. 
The HAPG crystal was placed 825 mm from the gas jet (X-ray source), and the X-ray CCD is also 825 mm from the crystal to maintain a 1-to-1 focus with the cylindrically curved crystal (required for optimal spectral resolution when using mosaic crystals such as HAPG~\cite{Zastrau2013}).
The crystal was tilted 11.8$^\circ$ from the laser and X-ray axis to observe X-rays in the 9 keV region.
A single layer of aluminium foil ($10\pm2~\rm{\mu m}$) was placed in front of the crystal at 45$^\circ$, to further protect the crystal from laser damage in the case of the tape drive failing.
The X-ray CCD was an Andor DX435, with a $2048\times1024$ chip with $13~\mathrm{\mu m}$ pixels, covered by an additional $10\pm2~\mathrm{\mu m}$ aluminium foil layer to prevent light leaks.
The CCD is cooled to $5^o$C and readout noise is on order of 1 count per spectral pixel, or $\approx0.1\%$ of a single photon.
The combined transmission of the Kapton tape and aluminium foils is 75\% at 9 keV.\\ 

\textbf{Broadband spectrum fitting.}
The spectral shape of the broadband on-axis X-ray beam was assumed to be synchrotron-like, characterised by a critical energy, $E_{\mathrm{crit}}$, and following the definition of Jackson~\cite{Jackson}.
That is, 
\begin{equation}
\frac{d^{2}I}{dEd\Omega}\propto(E/E_{\mathrm{crit}})^{2}\kappa^{2}_{2/3}[E/2E_{\mathrm{crit}}]
\end{equation}

where $\kappa_{2/3}[x]$ is a modified Bessel function of order 2/3. \\

\textbf{XAS processing.} The normalised XAS absorption is calculated by fitting below and above the copper K-edge, subtracting the former, before dividing by the latter.
The process is a standard procedure, and we follow the method outlined in our previous work~\cite{Kettle2019}, and that of Cook and Sayers~\cite{CookSayers1981}.
For the EXAFS processing, a spline fit is made to the general profile edge shape, which represents the idealised absorption of a bare atom, i.e. without local scattering of the outgoing photo-electron.
Subtracting it we are left with the oscillations of the EXAFS features, $\chi(k)$, the EXAFS fine-structure function, where $k=\sqrt{2m_e(E-E_0)/\hbar^2}$ is X-ray wavenumber and $E_0$ is the absorption threshold energy. 
$\chi(k)$ is weighted by a factor of $k^2$ which is a standard approach, chosen to highlight oscillations further from the edge.
Note that while our spectral resolution results in a small amount of broadening on the fine XANES features, there is no meaningful effect on the broader EXAFS structure. 
The next step in the processing is to Fourier transform to $\chi(R)$.
The software \textit{Larch}~\cite{Newville2013} was used to perform the spline fitting and subsequent Fourier transform to $R$-space.

\section*{Data availability}
The data that support the findings of this study are openly available~\cite{OurData}.
All other data are available from the corresponding author (or other sources, as applicable) on reasonable request. 


\bibliography{references} 
\bibliographystyle{naturemag}

\section*{Acknowledgements}
We wish to acknowledge the support of the staff at the Central Laser Facility.
This project has received funding from the European Research Council (ERC) under  European Union's Horizon 2020 research and innovation programme (gant no 682399), the John Adams Institute for Accelerator Science (STFC: ST/P002021/1, ST/V001639/1), the EPSRC (EP/S001379/1, EP/V044397/1, EP/N027175/1), the US DOE grant No. DE-SC0022109, the Knut and Alice Wallenberg Foundation (KAW 2019.0318), the Swedish Research Council (2019-04784), and the Helmholtz Association (VH-NG-1338).

\section*{Author Contributions}

B.K., C.C., E.E.L., E.G., M.J.V.S., F.A., N.C., K.F., O.L., P.P.R, D.R., S.J.R., G.S., K.S., D.R.S., M.S., A.G.R.T., C.T. and S.P.D.M. contributed to the planning and execution of the experiment.
S.A. and C.S. contributed to the experiment targetry.
R.A.B. and R.W. provided simulations for the experimental setup.
T.I.H. provided the reference synchrotron absorption scans. 
B.K., C.C., E.E.L., E.G., M.J.V.S., K.S., M.S. and S.P.D.M. performed analysis of the experiment data.
B.K. wrote the paper with contributions from C.C., E.E.L., E.G., M.J.V.S., K.F., D.R.S., O.L., M.S., A.G.R.T. and S.P.D.M.

\section*{Competing Interests}
The authors declare no competing interests.

\end{document}